\documentclass[twocolumn,aps,showpacs]{revtex4}
\usepackage{graphicx}
\usepackage{dcolumn}
\usepackage{color}
\usepackage{amsmath}
\usepackage{bm}
\def\prl#1#2#3{{ Phys. Rev. Lett.} {\bf #1}, #2 (#3)}

\def\pre#1#2#3{{ Phys. Rev. E} {\bf #1}, #2 (#3)}

\def\epl#1#2#3{{ Europhys. Letts.} {\bf #1}, #2 (#3)}

\def\eg{e.g.~}

\def\bc{\begin{center}}
\def\ec{\end{center}}
\def\eqn{\end{equation}\noindent}
\def\eqnr{\end{eqnarray}\noindent}
\def\beqr{\begin{eqnarray}}
 
\begin{document}
\title{Relaying phase synchrony in chaotic oscillator chains}
\author{Manish Agrawal$^{1}$, Awadhesh Prasad$^{1}$, and Ram Ramaswamy$^{2,}$\footnote{Present address:  University of Hyderabad, Hyderabad 500 046, India}}
\affiliation{$^{1}$\textit{Department of Physics and Astrophysics, University of Delhi, Delhi 110007,India}\\
$^{2}$\textit{School of Physical Sciences, Jawaharlal Nehru University, New Delhi 110067, India}}

\begin{abstract}
We study the manner in which the effect of an external drive is transmitted 
through mutually coupled response systems by examining the phase synchrony 
between the drive and the response. Two different coupling schemes are used. 
Homogeneous couplings are via the same variables, while heterogeneous
couplings are through different variables. With the latter scenario, synchronization 
regimes are truncated with increasing number of mutually coupled oscillators,
in contrast to homogeneous coupling schemes. Our results are illustrated 
for systems of  coupled chaotic  R\"ossler oscillators. 
\end{abstract}

\maketitle

\section{Introduction}
The many flavours of synchrony that emerge as a consequence of different coupling scenarios have
been examined in detail over the last couple of decades. The roles of  topology and  
nonlinearity have been studied, and a fair understanding of the different
forms of synchrony that can arise---given a specific scheme through which dynamical systems  
interact with each other---is understood to some degree \cite{pecora}. Synchronization in various forms is common in forced and coupled nonlinear systems \cite{pikovsky1, general, fuzisaka, pikovsky2, rulkov, rosen1}. The manner in which the interacting subsystems are coupled plays a crucial role in determining which form of synchrony arises. For example, when identical 
nonlinear systems are coupled unidirectionally with one subsystem (the master)  driving the response subsystem (the slave) complete synchronization occurs. When the coupled systems are not identical, generalized synchronization can occur  \cite{rulkov, kocarev}. ``Mixed'' synchrony is observed in counter--rotating coupled oscillators \cite{chaos} while phase synchronization occurs in mutually coupled chaotic systems: the phases of interacting systems are entrained while the amplitudes remain uncorrelated \cite{rosen1}. This is a phenomenon of great interest due to potential applications in different fields, ranging from physics, chemistry to biological and medical sciences \cite{disciplines}. 

Since the coupling can be uni-- or bi--directional and can be linear or nonlinear \cite{dhamala}, and can be through similar or dissimilar variables \cite{rajat1, rajat2}, as the number of interacting components increases, the possible variations grow exponentially. Our interest in  present work is to examine a small number of ``typical'' patterns or motifs of coupling, and investigate the different patterns of synchronization phenomena that result. An additional motivation arises from the fact that in a variety of natural systems that are subject to forcing, the modulation can be either direct, namely when a given system is itself subject to driving, or indirect, when it is coupled to another system
which is the one that is being modulated. Such indirect modulation is likely to be operative in  biological phenomena \cite{murray} or in networks of coupled dynamical systems \cite{network}.

In the present paper we study coupled chaotic oscillators which are externally forced. We find that as the number of mutually coupled oscillators increase, the phase synchronization (PS) regime gets truncated if the coupling is heterogeneous, namely through different variables, while the same does not hold if the coupling is homogeneous, namely  through the same variables. 

The different coupling schemes are discussed in the following Section II where we also study phase synchronization between the drive and the response. The effect  of the drive is further examined  in Section III. The measure we used to determine phase synchronization is based upon the variation in phase difference with time: this is discussed in an appendix that follows the concluding Section IV which presents a discussion and summary. 

\section{The Coupling Patterns}

We first consider the model system of three oscillators  coupled as schematically shown in Fig. 1.  Oscillators $O_2$ and $O_3$ (denoted by the subscripts in the variables) are diffusively and mutually coupled, and are driven by oscillator $O_1$.
The response oscillators $O_2$ and $O_3$ are identical and are distinct from oscillator $O_1$, there is a parameter mismatch in the frequencies. The scheme in Fig. 1(a) is termed  ``heterogeneous" since the driving is effected through a variable that is not involved in the coupling, while that in Fig. 1(b) is termed ``homogeneous" since the driven and coupled variables are the same.   
\begin{figure}
\includegraphics [scale=0.55]{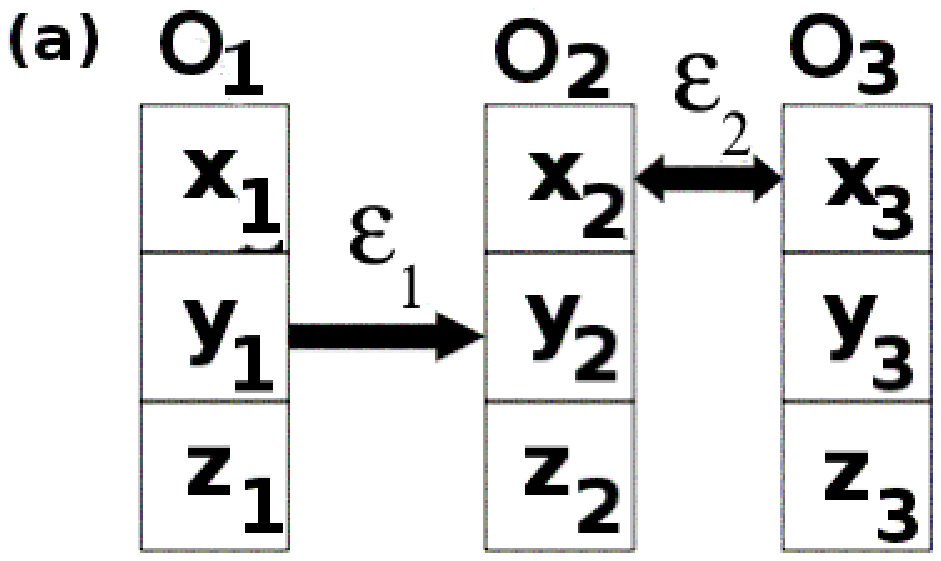}
\includegraphics [scale=0.55]{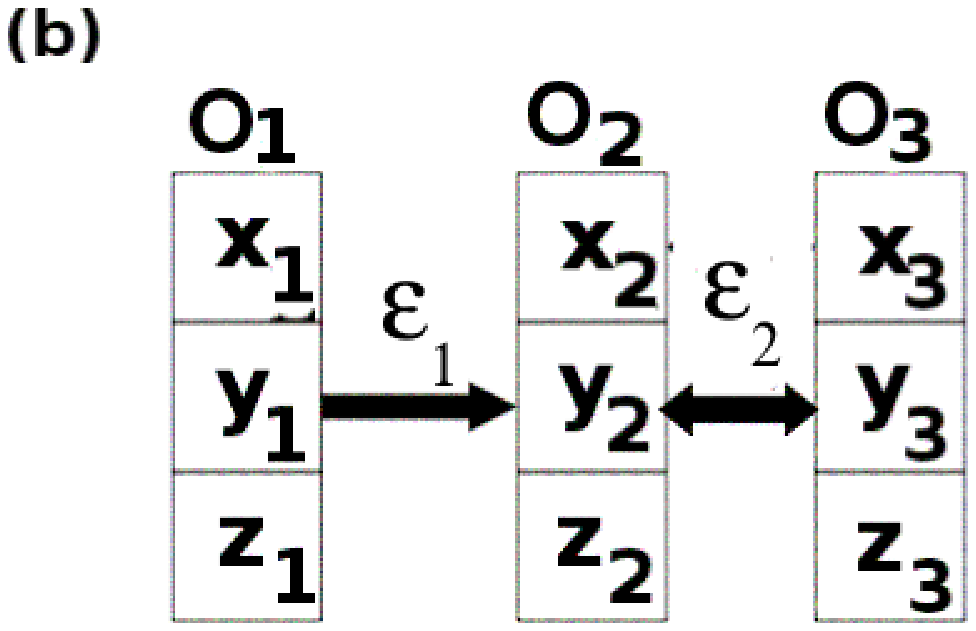}
\caption{(a) Heterogeneous coupling: the (nonidentical) chaotic oscillator $O_1$  is coupled to two mutually coupled identical oscillators through one of the variables, here $y$, while the identical oscillators $O_2$ and $O_3$  are symmetrically and bidirectionally coupled through the $x$ variable. (b) Homogeneous coupling: all the interactions are through the same variable, here $y$.}  
\label{fig:model1}
\end{figure}

\subsection{Heterogeneous Coupling}
Consider the system of three coupled R\"ossler chaotic attractors,
\begin{eqnarray}
\label{eq:ross}
\dot {x}{_1}&=&-y_1-z_1 \nonumber \\
\dot {y}{_1}&=&x_1+a_1 y_1 \nonumber \\
\dot {z}{_1}&=& b_1+z_1 (x_1-c_1)\nonumber \\
\ ~ \nonumber\\
\dot {x}{_2}&=&-y_2-z_2+\epsilon_2 (x_3-x_2) \nonumber \\
\dot {y}{_2}&=&x_2+a_2 y_2+ \epsilon_1 (y_1-y_2) \nonumber \\
\dot {z}{_2}&=& b_2+z_2 (x_2-c_2)\nonumber \\
\ ~ \nonumber\\
\dot {x}{_3}&=&-y_3-z_3+\epsilon_2 (x_2-x_3) \nonumber \\
\dot {y}{_3}&=&x_3+a_2 y_3 \nonumber \\
\dot {z}{_3}&=& b_2+z_3(x_3-c_2).
\end{eqnarray}
\noindent

We study the 
different synchronization states in coupling parameter's space, $\epsilon_1$ and $\epsilon_2$.
The internal parameters of these oscillators are fixed at $a_1$=$0.2$, $b_1$=$0.2$, and $c_1$=$5.7$
 (for driving oscillator $O_1$ );
while $a_2$=$0.15$, $b_2$=$0.2$, and $c_2$=$10$ (for mutually coupled oscillators, $O_2$ and $O_3$).  At these set of parameters all oscillators show chaotic motion.
The natural frequencies $f_i$ turn out to be $f_1=1.079$ and $f_2=f_3=1.041$ respectively.

\begin{figure}
\includegraphics [scale=0.325]{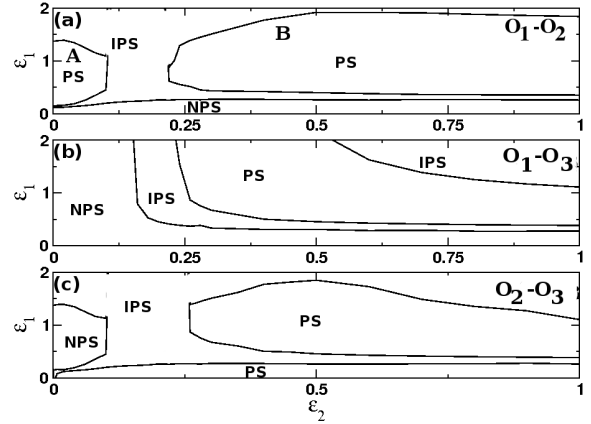}
\caption{Schematic phase diagram for the phase synchronization states as a 
function of  $\epsilon_1$ and $\epsilon_2$ between (a) oscillator $O_1$ and the directly forced response  oscillator $O_2$, (b) $O_1$ and the indirectly  driven oscillator $O_3$, and
(c) the mutually coupled driven oscillators $O_2$ and $O_3$. The symbols PS, IPS, and NPS 
represent the phase synchronization, imperfect phase synchronization, and
unsynchronized states respectively.}  
\label{fig:param}
\end{figure}   

Shown in Fig. \ref{fig:param} are schematic phase diagrams  with regard to phase synchrony as a function of $\epsilon_1-\epsilon_2$. The different regimes are characterized through a measure based upon the time--dependence of the phase difference (see Appendix A). 
In order to verify the phase synchrony  \cite{rosen1,rosen2} we use the phase for R\"ossler oscillators in Eq. (\ref{eq:ross}) \cite{pikovsky3,gorya}, namely $\phi{_i} = \arctan 
(y_i/x_i),~~ i$=1, 2, 3. Phase synchronization (PS) occurs when the phase difference between two
 interacting oscillators $\vert \phi_i-\phi_j\vert$ saturates \cite{rosen2} (see Fig. \ref{fig:param} for details). In imperfect phase synchrony (IPS) regions in 
 Fig. \ref{fig:param}, the subsystems are phase locked but are subject to occasional slips.
The value of  $\vert\phi_i-\phi_j\vert$ varies during the evolution of the chaotic system 
\cite{park}, and it changes in a step-wise manner.
Each step corresponds to a phase synchronized state under a particular phase locking condition.
The jump between consecutive steps occurs in multiples of $\pi$ \cite{park}. In quasiperiodically forced systems \cite{agrawal} the phase differences in the IPS state also change in arbitrary multiples of $\pi$.  Fig. \ref{fig:param} shows phase synchronization states between (a) the driving oscillator $O_1$ and directly driven oscillator $O_2$, (b) the driving oscillator $O_1$ and indirectly driven oscillator $O_3$, and (c) mutually coupled response oscillators $O_2$ and $O_3$.

Transitions among the phase synchronization states when the  coupling parameters are varied are depicted in Fig. \ref{fig:param}(a). In one case unsynchronized oscillators (NPS) become phase synchronized (PS, region B) via the imperfect phase synchronized state (IPS), and in another transition the phase synchronization (PS, region A) transitions to the IPS state. As shown in Fig. \ref{fig:param}(b) the indirectly driven oscillator $O_3$ is initially unsynchronized to the driving oscillator $O_1$, but with the increase of mutual interaction between $O_2$ and $O_3$ the oscillator $O_3$ becomes phase synchronized to the drive $O_1$.
This phase synchronization behavior of $O_3$ clearly signifies the transmission of drive to the mutually coupled oscillators. Phase synchronization between the response subsystems $O_2$ and $O_3$  is seen in Fig. \ref{fig:param} (c). By comparing the Figs. \ref{fig:param}(a), \ref{fig:param}(b), and \ref{fig:param}(c) we observe that mutually coupled oscillators $O_2$ and $O_3$ are phase synchronized when each of the driven oscillators phase synchronizes with the drive. 
   
\begin{figure}
\includegraphics [scale=0.325]{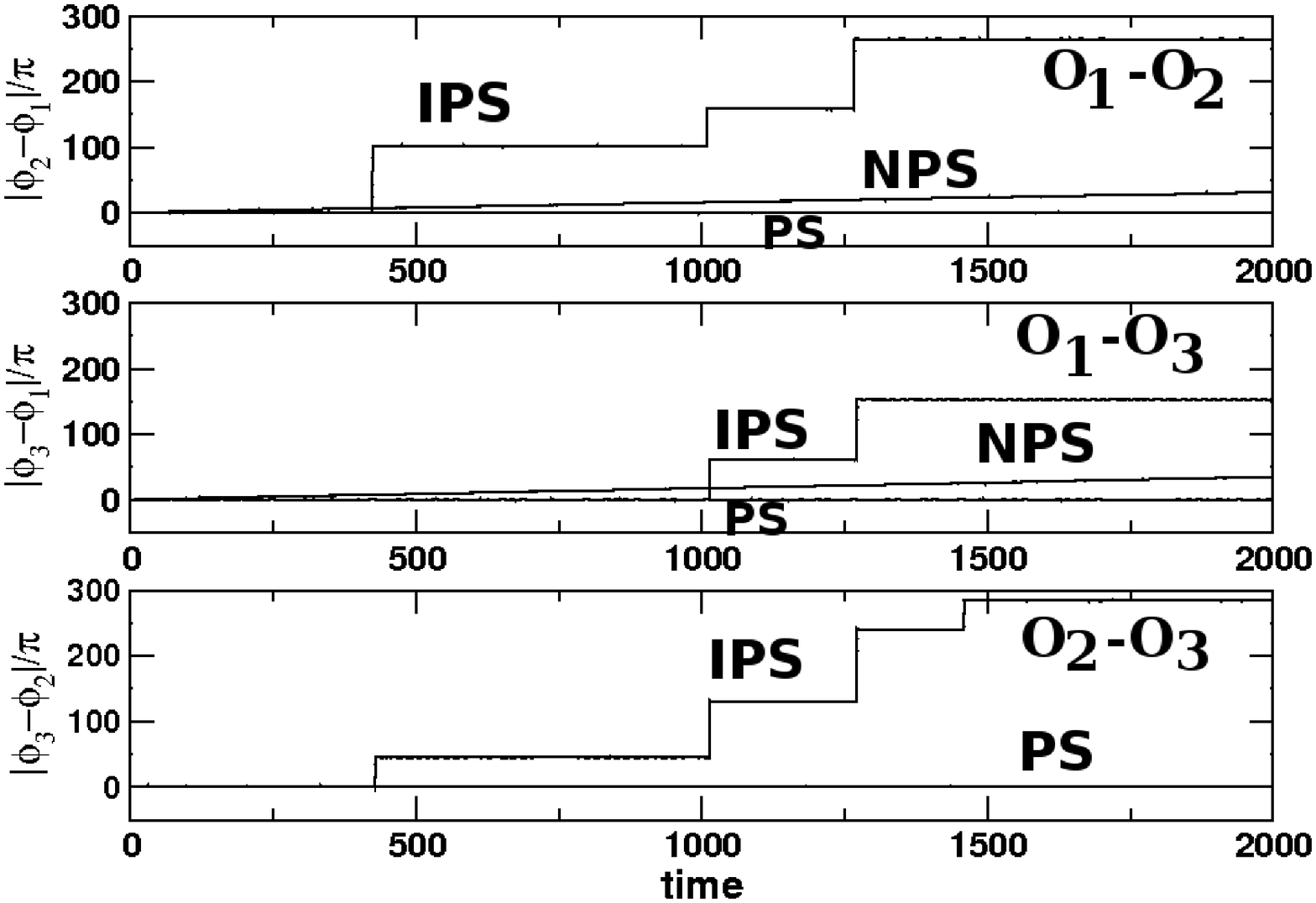}
\caption{Phase differences between (a) oscillators $O_1$ and $O_2$ for PS at $\epsilon_1$ = 1, IPS
at $\epsilon_1$=2, and NPS at $\epsilon_1$=0.15;  
(b) oscillators $O_1$ and $O_3$ for  PS at $\epsilon_1$=1, IPS at $\epsilon_1$=1.2, and NPS at
  $\epsilon_1$=0.2; and
(b) oscillators $O_2$ and $O_3$ for PS at $\epsilon_1$=1 and IPS at $\epsilon_1$=1.5. The bidirectional coupling parameter is fixed at $\epsilon_2$=1 in all cases.}
\label{fig:phase}
\end{figure}  

The above results  are further illustrated at different points in Fig. \ref{fig:phase}. Fig. \ref{fig:phase} (a) describes oscillators $O_1$ and $O_2$, Fig. \ref{fig:phase} (b) describes the behavior of  $O_1$ and $O_3$, while the phase synchronization states in between the mutually coupled oscillators $O_2$ and $O_3$ is shown in Fig. \ref{fig:phase} (c). The bidirectional coupling parameter $\epsilon_2$ is kept fixed at $\epsilon_2$=$1$, and the different curves are drawn for different values of the coupling parameter $\epsilon_1$. These curves clearly show the behavior of interacting subsystems:  the phase difference remains bounded for PS whereas it continuously grows with time in NPS. 

\begin{figure}
\includegraphics [scale=0.325]{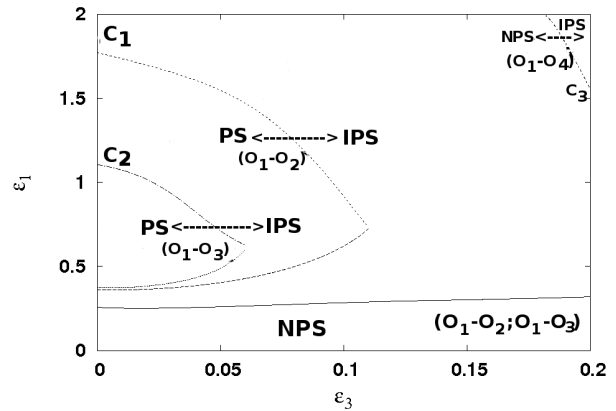}
\caption{Phase diagram (schematic) for the coupling parameters $\epsilon_3$ and $\epsilon_1$. $\epsilon_3$ presents the coupling between $O_3$ and $O_4$. The coupling between $O_2$ and $O_3$ is kept fix at $\epsilon_2$=$1$. $C_1$, $C_2$, and $C_3$ are separating curves for the pairs $O_{1}$-$O_{2}$, $O_{1}$-$O_{3}$, and $O_{1}$-$O_{4}$ respectively. These curves separate PS, IPS, and NPS for respective pairs. Details are given in text.}
\label{fig:minimal}
\end{figure} 

We further consider a chain of $N$ oscillators with nearest--neighbor coupling. Phase synchronization between the drive $O_1$ and responses ($O_i$, $i=2, 3,\ldots, N$) is lost in such systems as we increase the number of response
oscillators. To determine whether the loss of phase synchrony is abrupt or gradual, we first study the case of an extended system consisting of a driving oscillator $O_1$ and \textit{three} mutually coupled response oscillators (\textbf{$O_1$}$\rightarrow$\textbf{$O_2$}$\leftrightarrow$\textbf{$O_3$}$\leftrightarrow$\textbf{$O_4$}). Among three mutually coupled subsystems, $O_2$ and $O_3$ are connected by $\epsilon_2$ while subsystems $O_3$ and $O_4$ are connected by the coupling parameter $\epsilon_3$. Fig. \ref{fig:minimal} shows the phase synchronization state between the subsystem $O_1$ and driven subsystems $O_2$, $O_3$, and $O_4$. To observe the effect of forcing in extended systems, the coupling between $O_2$ and $O_3$ is fixed at $\epsilon_2$=$1$, while Fig.~\ref{fig:minimal} is for varying  $\epsilon_3$ and $\epsilon_1$.  

Our numerical results (see Fig. \ref{fig:minimal}) suggest that phase synchronization is not lost abruptly: with the increase of coupling between $O_3$ and $O_4$ the single bounded region of phase synchronization shrinks between $O_{1}$-$O_{2}$, $O_{1}$-$O_{3}$ and the
drive and response subsystems turns out to be imperfect phase synchronized (IPS) and then phase unsynchronized (NPS).
The phase synchronization states in Fig. \ref{fig:minimal} are observed for all possible drive-response pairs 
($O_{1}-O_{2}$, $O_{1}-O_{3}$, and $O_{1}-O_{4}$), and in comparison of first two pairs we find that the phase space area for phase 
synchronization (PS) is larger in $O_{1}-O_{2}$ then in $O_{1}-O_{3}$ case. The larger phase space area for 
PS in $O_{1}-O_{2}$ than in $O_{1}-O_{3}$ shows that  in a chain of mutually coupled oscillators, the ability of the drive to synchronize the system decreases with distance from the drive.

It should be noted that curves $C_1$ and $C_2$ are boundaries between the phase synchronized (PS) and imperfect phase synchronized (IPS) region for $O_{1}-O_{2}$ and $O_{1}-O_{3}$ pairs of subsystems respectively (see Appendix A). $C_3$ separates the phase unsynchronized state (NPS) from the IPS state for the $O_1-O_4$ pair. The decreasing ability  of the drive to cause synchrony is further verified in  Fig. \ref{fig:minimal} where we see that the oscillators $O_{1}$ and $O_{4}$ are largely phase unsynchronized (NPS) in given parameter range, though we observe the small appearance of IPS state in higher parameter values. The oscillators $O_{1}$ and $O_{4}$ continue to be in this IPS state for higher values of the varying parameters ($\epsilon_3$, $\epsilon_1$). 

\subsection{Homogeneous Coupling}
In this scheme two mutually coupled identical chaotic systems  are coupled unidirectionally to another (but nonidentical) chaotic oscillator, unidirectionally through the same variable (here $y$). See Fig. \ref{fig:model1}(b). The  equations of motion for three coupled R\"ossler oscillators in this scheme
are
\begin{eqnarray}
\label{eq:ross1}
\dot {x}{_1}&=&-y_1-z_1 \nonumber \\
\dot {y}{_1}&=&x_1+a_1 y_1 \nonumber \\
\dot {z}{_1}&=& b_1+z_1 (x_1-c_1)\nonumber \\
\ ~ \nonumber\\
\dot {x}{_2}&=&-y_2-z_2 \nonumber \\
\dot {y}{_2}&=&x_2+a_2 y_2+\epsilon_1 (y_1-y_2)+\epsilon_2 (y_3-y_2) \nonumber \\
\dot {z}{_2}&=& b_2+z_2 (x_2-c_2)\nonumber \\
\ ~ \nonumber\\
\dot {x}{_3}&=&-y_3-z_3 \nonumber \\
\dot {y}{_3}&=&x_3+a_2 y_3+\epsilon_2 (y_2-y_3) \nonumber \\
\dot {z}{_3}&=& b_2+z_3(x_3-c_2).
\end{eqnarray}
\noindent

\begin{figure}
\includegraphics [scale=0.325]{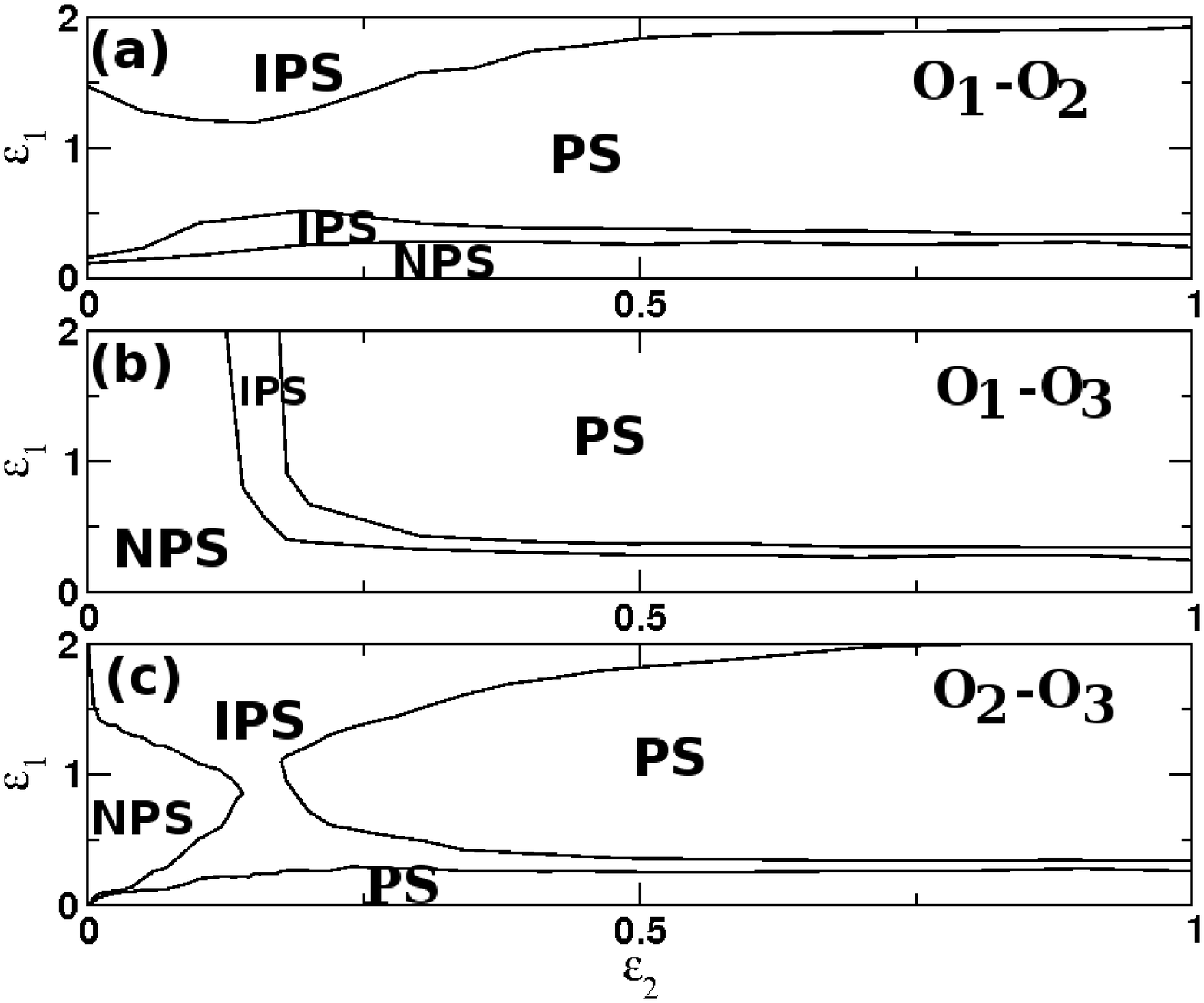}
\caption{Schematic phase diagram for the synchronization states as a function of the coupling 
parameters $\epsilon_1$ and $\epsilon_2$ in the homogeneous coupling scheme between (a) oscillator $O_1$ and the directly 
driven oscillator $O_2$, (b) $O_1$ and the indirectly driven oscillator $O_3$, and (c) 
mutually coupled oscillators $O_2$ and $O_3$. The symbols PS, IPS, and NPS have the same meaning as in Fig. \ref{fig:param}.} 
\label{fig:homo1}
\end{figure}

A schematic phase diagrams as a function of the parameters $\epsilon_1$  and $\epsilon_2$ is shown in Fig. \ref{fig:homo1}. Fig. \ref{fig:homo1} (a) shows the phase synchronization states
between the driving oscillator $O_1$ and directly driven oscillator $O_2$, while Fig. \ref{fig:homo1} (b)  shows the phase synchronization states between the driving oscillator $O_1$ and indirectly driven  oscillator $O_3$. The phase synchronization states between the mutually coupled response subsystems $O_2$ and $O_3$ are shown in Fig. \ref{fig:homo1} (c).

\begin{figure}
\includegraphics [scale=0.325]{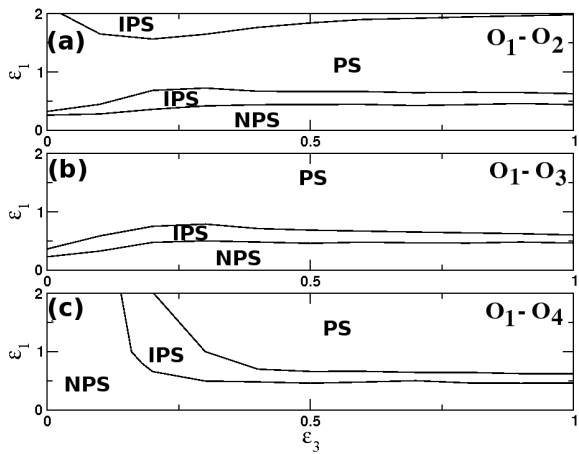}
\caption{Schematic phase diagram for the extended case, showing  synchronization states as a function of coupling parameters $\epsilon_3$ and $\epsilon_1$. $\epsilon_3$ is the coupling between $O_3$ and $O_4$, $\epsilon_1$
couples the drive to the responses. Phase synchronization states between 
(a) drive $O_1$ and the directly driven oscillator $O_2$ (b) $O_1$ and the indirectly driven oscillator $O_3$, and (c) $O_1$ and the indirectly driven oscillator $O_4$. In all cases, $\epsilon_2$= 1.} 
\label{fig:homo2}
\end{figure}

The occurrence of phase synchronization states between $O_1$ and  $O_2$ or $O_3$ is a consequence of the transmission of forcing through mutually coupled chaotic oscillators. In order to compare this effect with that of the
heterogeneous case (previous subsection) we consider the larger number of oscillators in a chain. As shown in Fig. \ref{fig:homo2} by increasing the number of oscillators  to 4 ($O_1 \rightarrow O_2 \leftrightarrow  O_3 \leftrightarrow  O_4$) with the coupling between $O_2$ and $O_3$ fixed at $\epsilon_2$ = 1 we find that the external drive transmission decreases with distance but the phase synchronization between the drive $O_1$ and responses $O_2$, $O_3$, and $O_4$  continues to occur so long as the coupling is heterogeneous. These results hold for even larger numbers of  mutually coupled oscillators.  A second difference between hetro- and homogeneous schemes is that there are multiple regimes of PS in the former case (see Fig. \ref{fig:param}) but only a single regime for latter.

\section{Forcing through mediating systems}
Natural systems are often modulated indirectly. Consider a drive-response pair  mediated by a number of mutually coupled  subsystems. In order to clearly understand the present case of external forcing via intermediate subsystems, the model system is crafted from both heterogeneous and  homogeneous coupling schemes. 
(Recall that the difference between these two coupling schemes is as follows: homogeneous coupling has both the drive and response subsystems coupled by the same variable, while heterogeneous coupling uses different variables.) 

\begin{figure}
\includegraphics[scale=0.4]{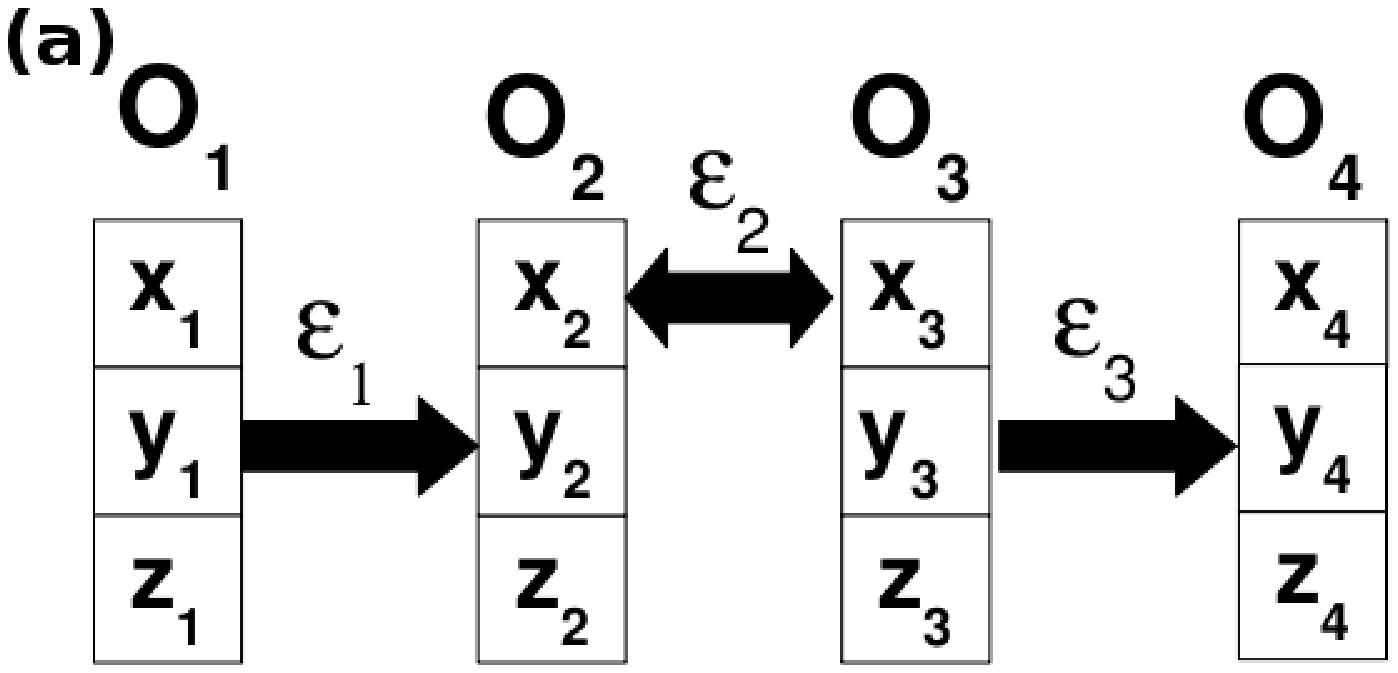}
\includegraphics[scale=0.6]{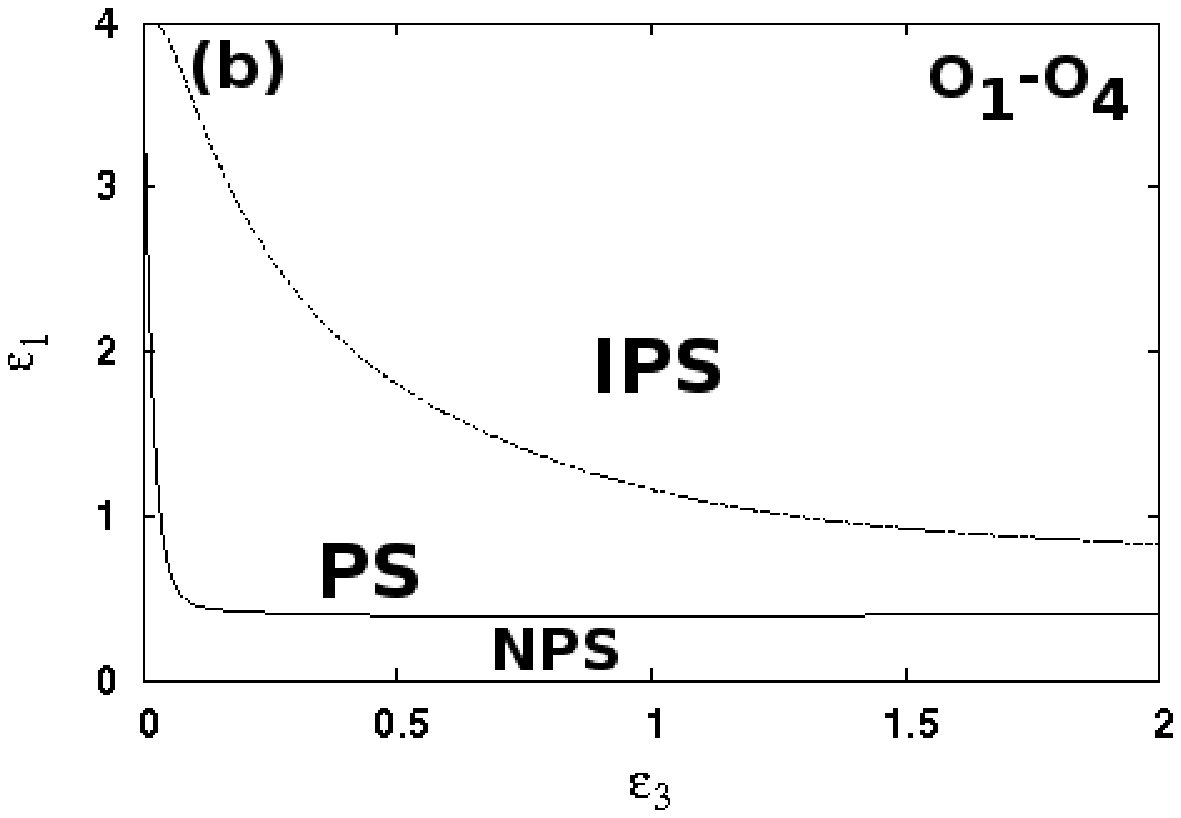}
\includegraphics[scale=0.4]{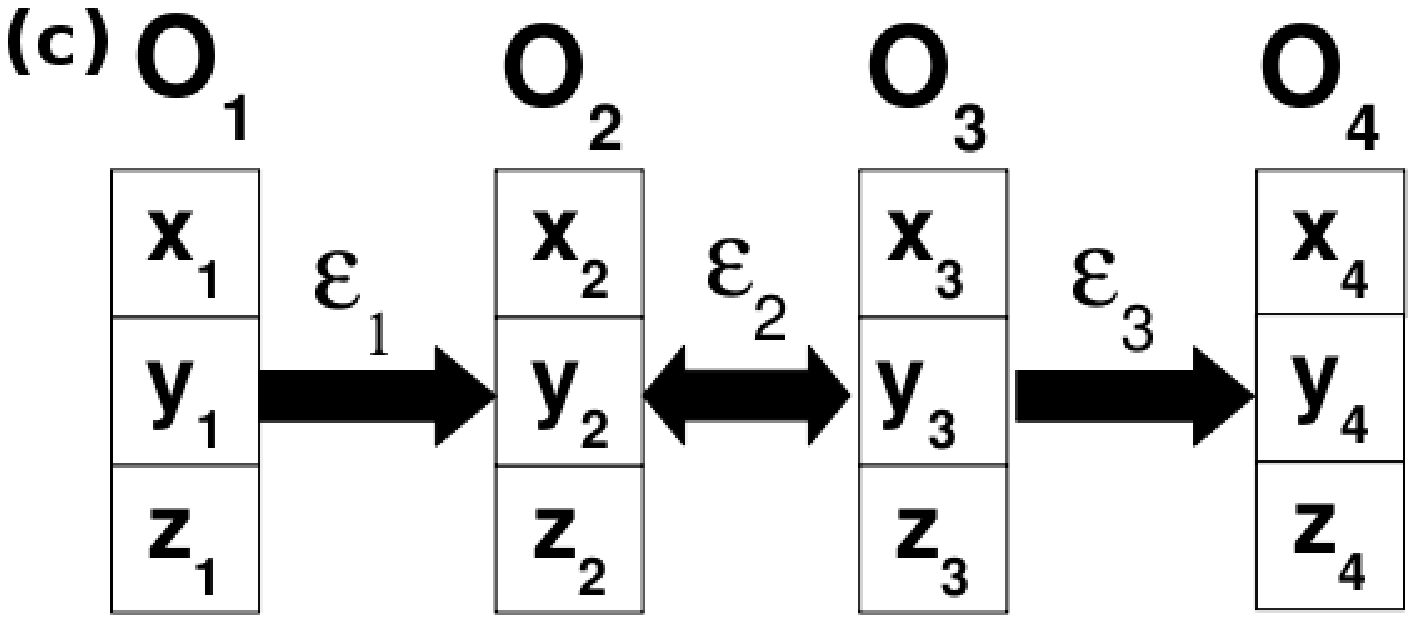}
\includegraphics[scale=0.6]{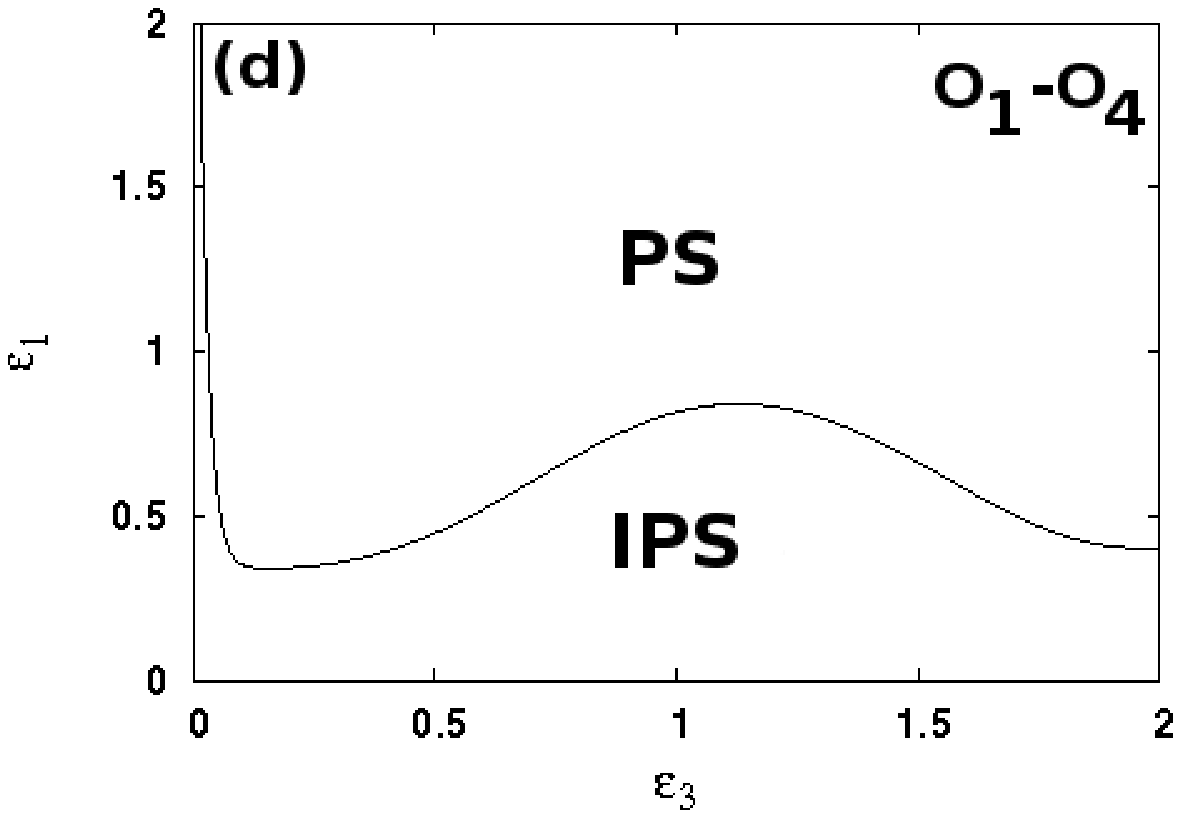}
\caption{Heterogeneous coupling: (a) the system of Fig. \ref{fig:model1} (a) drives a dissimilar oscillator $O_{4}$ through the response oscillator $O_{3}$, (b) a schematic phase diagram as a function of $\epsilon_{3}$ and $\epsilon_{1}$,  between oscillators $O_{1}$ and $O_{4}$ for fixed $\epsilon_{2}$ = 1. Homogeneous coupling: (c) all oscillators are connected through the same 
variable, here $y$, and (d) phase synchronization states between oscillators $O_{1}$ and $O_{4}$ for fixed $\epsilon_{2}$ = 1.}  
\label{fig:dforcing}
\end{figure}

The heterogeneous coupling is shown in
Fig. \ref{fig:dforcing} (a) where oscillator $O_{1}$  drives $O_2$ 
and $O_3$ which then drive oscillator $O_{4}$. If $\epsilon_3=0$ this reduces to the case of  Fig. \ref{fig:model1}(a) but for finite $\epsilon_3$ different phase synchronization states (PS, IPS, and NPS) are observed between oscillators $O_1$ and $O_4$. Results are shown in Fig. \ref{fig:dforcing} (b) for a representative value of $\epsilon_2=1$. 
As  the number of mutually coupled oscillators between the concerned drive-response pairs (\eg $O_1-O_4$) is increased, the PS regime is lost. This result though not presented here, is in consonance with the result of Fig. \ref{fig:minimal} for the heterogeneous coupling case. 

With homogeneous coupling, as shown in Fig. \ref{fig:dforcing} (c), phase synchrony between the terminal oscillators $O_1$ and $O_4$ is also achieved. Results are shown in Fig. \ref{fig:dforcing} (d) and it can be seen that for a given value of $\epsilon_{3}$ oscillators $O_1$ and $O_4$ become phase synchronized beyond a critical value of coupling strength $\epsilon_{1}$. The non-monotonic form of the border between IPS and PS in Fig. \ref{fig:dforcing} (d) is possibly due to two indirect forcing: $O_1$ is forcing $O_2$ and $O_3$, while the combined oscillators $ O_1, O_2$, and $O_3$ are forcing $O_4$. This PS regime persists even in case of increased number of mutually coupled intermediate oscillators. 
 
Heterogeneous and homogeneous coupling schemes thus clearly exhibit distinct dynamics. In the former, synchronization regimes are truncated while in the later case synchrony persists even when the number of mutually coupled oscillators is increased. The conjugate variables employed in heterogeneous coupling appear to provide an effective time-delayed interaction \cite{rajat1, packard}. 

By computing the conditional Lyapunov exponents \cite{parlitz}  as a function of  the external coupling strength,
we find that in general, phase synchronization  is not obtained when the drive  couples to $z$, so 
in the present studies, we  drive the  variable $y$ (see Eqns. (\ref{eq:ross}) and (\ref{eq:ross1})).
 Results are similar when drive is coupled with the $x$ variable as well.  

\section{Summary}

In the present work we have studied the effect of an external
drive on symmetrically and diffusively coupled response oscillators. 
Two different coupling schemes have been examined; these coupling schemes were based on the use of the two coupling parameters either on the same
variables (homogeneous coupling) of interacting oscillators or on different variables (heterogeneous coupling). Numerous combinations are possible, and this study is an attempt to consider some representative examples. 

The relaying of  the drive to both the directly and the indirectly forced subsystems is seen through the occurrence of phase synchronization.  For heterogeneous coupling phase synchronization between drive $O_1$ and responses $O_i$, $i=2,3,...,N$ is lost in extended systems, but in a gradual manner. Thus the transmission effect of external drive decreases sequentially in an array of mutually coupled response oscillators. In case of homogeneous coupling the external drive is transmitted to each of the mutually coupled response subsystems. The effect of indirect external forcing has been also observed for a typical model with mediating oscillators. 

The occurrence of phase synchrony is of particular significance due to potential applications in diverse fields such as biological systems \cite{murray}, networks of coupled dynamical systems \cite{network} among others. The present results indicate that the transmission of drive to an array of mutually coupled response oscillators depends strongly on the coupling pattern, and it is very likely that the topology of coupling will also play a significant role. A study of different coupling motifs in order to determine the manner in which the drive is transmitted in more complex networks is therefore presently under way \cite{agrawal2}.

\begin{acknowledgments}
MA is grateful to the University Grants Commission for the RFSMS fellowship, and AP \& RR would like to thank the Department of Science and Technology India, for financial support.
\end{acknowledgments}

\appendix
\section{Phase Synchronization}

The manner in which the phase difference varies with time distinguishes the states of PS, IPS, and NPS.  When two interacting systems are phase synchronized, the phase difference between the systems remain bounded \cite {rosen1} while the phase difference between the interacting subsystems continuously grows when the systems are phase unsynchronized (NPS). In case of imperfect phase synchronization the phase difference grows in multiples of $\pi$ \cite{park, agrawal}. This variation in phase difference is used to deduce a measure for the identification of different phase synchronization states.  The averaged phase difference increment $\Phi$
is the time average of the derivative of the instantaneous phase difference $\Delta \phi_t = \mid \phi_2-\phi_1 \mid$  at time $t$, namely 
\begin{eqnarray}
\label{eq:measure2}
\Phi &=& \langle \Delta \phi_{t+1}- \Delta \phi_{t} \rangle .
\end{eqnarray}
\noindent

\begin{figure}
\includegraphics [scale=0.325]{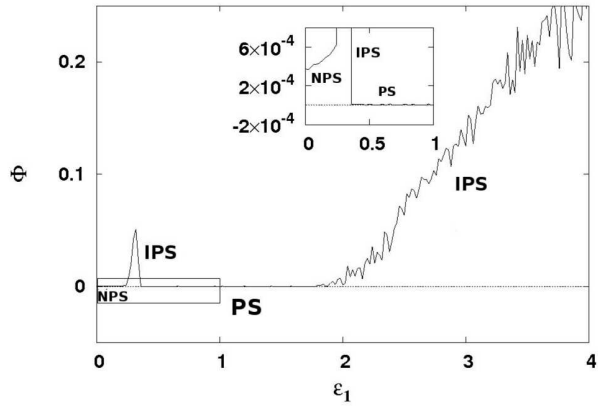}
\caption{Phase synchronization measure for \textit{heterogeneous} coupling with $\epsilon_2$ = 1,  varying $\epsilon_1$. The inset shows the order of the averaged phase difference increment  in the three different states.} 
\label{fig:measure}
\end{figure}

\begin{enumerate}
\item If the variation of phase difference remains bounded between $0$ and $\pi$, $\Phi$ should be close to zero in case of phase synchronization. 

\item For IPS the phase difference between the two  interacting subsystems grows in multiples of $\pi$; this results in large values of $\Phi$.

\item When the interacting subsystems are out of phase synchrony, the phase difference increases with time but  $\Phi$ in considerably lower than it is in the case of IPS.
\end{enumerate}
 
 Fig. \ref{fig:measure} shows the averaged phase difference increment $\Phi$ as a function of the  parameter $\epsilon_1$ for the system  in Fig.\ref{fig:model1} (a) with  fixed $\epsilon_2$ = 1. The inset shows the regions of PS, IPS, and NPS. This measure is qualitative in the sense that sharp boundaries between these states (Figs. \ref{fig:param}, \ref{fig:minimal}, \ref{fig:homo1}, \ref{fig:homo2}, \ref{fig:dforcing}(b) and (d)) cannot be drawn in a quantitative manner. We  use  thresholds  for $\Phi$ as 10$^{-5}$, 0.0001, and 0.1 to assign the boundary for the transition PS to IPS/NPS, NPS to IPS, and NPS to IPS respectively.

\end{document}